\documentclass[twocolumn,prl,aps,superbib,tightenlines,floatfix]{revtex4}
\usepackage{amsfonts}
\usepackage{amsmath}
\usepackage{amssymb}
\usepackage{graphicx}
\begin{document}
\bibliographystyle{apsrev}
\draft
\input epsf.sty \flushbottom
\title{Shot noise in parallel wires}
\author{Johan Lagerqvist}
\author{Yu-Chang Chen}
\author{Massimiliano Di Ventra\cite{MD}}
\affiliation{Department of Physics, University of California, San Diego, La Jolla, CA 92093-0319}
\begin{abstract}

We report first-principles calculations of shot noise properties of parallel carbon wires in the regime in which the interwire distance is much smaller than the inelastic mean free path. We find that, with increasing interwire distance, the current approaches rapidly a value close to twice the 
current of each wire, while the Fano factor, for the same distances, is still larger than the Fano factor of a single wire. This enhanced Fano factor is the signature of the 
correlation between electron waves traveling along the two different wires. In addition, we find that the 
Fano factor is very sensitive to bonding between the wires, and can vary by orders of magnitudes by changing the 
interwire spacing by less than 0.5~\AA.  While these findings confirm that shot noise 
is a very sensitive tool to probe electron transport properties in nanoscale conductors, they also suggest that 
a considerable experimental control of these structures is required to employ them 
in electronics. 

\end{abstract}
\maketitle

\section{Introduction}
In recent years we have witnessed an increased interest in the transport properties of molecules sandwiched  
between two bulk electrodes.~\cite{refs1,refs4,refs5,refs6,refs7,refs8,refs9,refs11,refs12,refs13,reed,refs14,refs2,refs3,refs10} This interest is partly motivated by the possible use of such structures in future electronic 
applications, and partly by the desire to understand electronic transport in atomic-scale conductors. 
In most of the experiments reported so far the number of molecules sandwiched between two bulk electrodes is unknown.~\cite{refs1,refs5,refs6,refs11,refs12,refs13,reed,refs14,refs4,refs7,refs8,refs9} Even in experiments where this number is supposedly very small~\cite{refs1,refs5,refs6,refs11,refs12,refs13,reed,refs14} 
it is still unclear whether a single molecule is responsible for the reported current-voltage (I-V) 
characteristics or if many molecules contribute.~\cite{refKirc1,refKirc2} Irrespective of this issue, future electronic applications of molecular wires will likely require assembly of many components with interwire separations at the nanometer scale. 
At these distances electrons are less likely to experience inelastic scattering events so that quantum interference between electron waves travelling in adjacent wires dominates. We can therefore expect that 
the transport properties of an ensemble of coupled wires are not necessarily a simple superposition of the transport properties of each isolated wire.~\cite{langavouris2000} Past work has indeed addressed this issue by studying the 
electric current in parallel atomic-scale wires as a function of their 
separation.~\cite{yalirakiratner98,magoga99,langavouris2000} For instance, Lang and Avouris have found 
large variations of the low-bias conductance of two carbon wires connected in parallel to bulk 
electrodes as a function of their separation.~\cite{langavouris2000} These variations have been attributed to 
both a direct bonding between the wires and an indirect interaction due to the presence of the metal 
electrodes.~\cite{langavouris2000} 

In this paper we want to build on the knowledge gained from these earlier studies 
and investigate a related transport property: steady-state current fluctuations (shot noise) in parallel wires. 
Our motivation is twofold: for one, current fluctuations need to be small for nanoscale components to have 
practical applications. Understanding the magnitude of these fluctuations in single wires and to what extent 
interference effects between coupled wires modify shot noise is therefore paramount to progress in nanoscale 
electronics. Secondly, shot noise contains more information on the transport properties of particles  
than the average current.\cite{buttiker} In the case of electrons, for instance, shot noise provides information 
on their energy distribution~\cite{bulashenko}, kinetics,~\cite{landauer} and
interactions due to the Coulomb repulsion and the Pauli exclusion
principle~\cite{liu}. Shot noise has also been found to depend on the atomic details of nanoscale wires like their 
length and coupling to bulk electrodes.~\cite{chendiventra2003} All these studies suggest that shot noise is a very 
sensitive tool to study electron transport properties in nanoscale conductors and we hope that our work will 
motivate further experimental work in this direction. 

To the best of our knowledge very few reports have appeared 
in literature on the shot noise properties of parallel conductors connected to bulk 
electrodes.~\cite{iannaccone97,gattobigio02,scholl03} 
All of them employ simple models of the electronic structure of the system and/or of the 
current correlations on each electrode.~\cite{iannaccone97,gattobigio02,scholl03} For instance, Iannaccone 
{\it et al.} have shown that if an electron transmitted from one wire in a given lead does not enter the 
adjacent wire on the same lead, then the shot noise of each wire adds classically to give a Fano 
factor equal to the Fano factor of one wire.~\cite{iannaccone97} We call this case an ``uncoupled-wire'' 
situation. However, when the distance between the wires is few nanometers it is unlikely that the 
above case can occur: any phase-breaking process (whether it is due to inelastic scattering, quasi-elastic 
dephasing or electron-electron scattering) would not be enough to break completely the correlation between 
waves traveling along the two distinct wires. The shot noise of the total system, therefore, should not be a simple classical summation of noise of each individual component. This is precisely the physical situation we want to 
investigate here. 

\begin{figure}
\includegraphics[width=.48\textwidth]{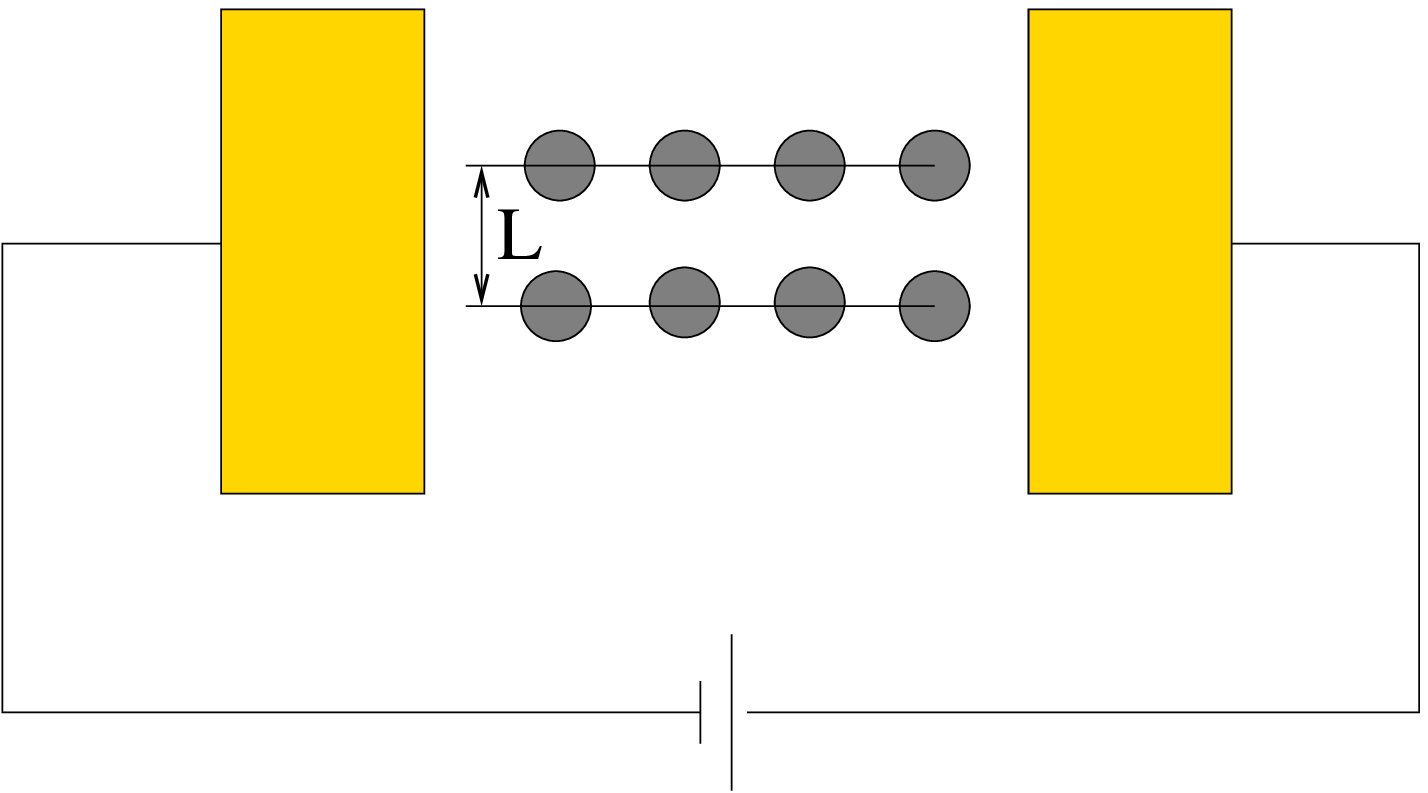}
\caption{Schematic of the system investigated. It consists of two parallel wires, separated by a distance L, between two biased electrodes.}
\label{fig1}
\end{figure}

Following the work of Lang and Avouris~\cite{langavouris2000} we study the transport properties 
of a system comprising two carbon wires of different lengths (also known as 
cumulene carbon atom chains) between two bulk electrodes (see schematic in Fig.~\ref{fig1}). 
The bulk electrodes are modelled with a semi-infinite 
uniform-background model (jellium model).~\cite{langt,diventralang021,diventralang022} The interior electron density of the
electrodes is taken close to the value for metallic gold ($r_{s}\approx3$). Note that in the work 
reported in Ref.~\onlinecite{langavouris2000} the 
interior electron density of the electrodes has been chosen to have $r_{s}\approx2$. 
Therefore, the actual shape of the 
conductance as a function of interwire distance is different than the one reported in the present work. However, the physical trends are similar. 

The cumulene-chain system is structurally and electronically 
quite simple, such that the first-principles 
calculations reported here can be done with a relatively high level of accuracy. In addition, cumulene atom chains 
can be loosely identified as the smallest ``carbon nanotubes'' possible. The results presented in this work 
can thus bear some relevance in nanotube electronics. We find, in agreement 
with previous reports,~\cite{langavouris2000} that the current varies with interwire distance and saturates 
to a value close to twice the current of each wire with increasing wire separation. On the other hand, 
the Fano factor can vary by orders of magnitudes by changing the 
interwire spacing by less than 0.5~\AA. Furthermore, when the current is very close to the value of twice 
the current of each wire, the Fano factor is still larger than the Fano factor of a single wire. 
This enhanced Fano factor is the signature of the 
correlation between electron waves traveling along the two different wires. At the distances considered here, this correlation is unlikely to be destroyed by inelastic scattering effects, therefore this effect should be observed 
in parallel nanoscale wires.

\section{Theoretical approach}

In the theory of mesoscopic systems, shot noise is generally expressed in terms of the transmission probabilities $T_n$ of 
each conducting channel $n$. Instead of discussing shot noise in a system that carries an average current $I$,  
it is customary to report on the ratio (called Fano factor) 
between the shot noise and its Poisson (uncorrelated electrons) limit $2eI$, 
where $e$ is the electron charge. For a two-terminal conductor the Fano
factor can then be expressed, in linear response, as 
\begin{equation}
F=\sum_{n}T_{n}\left(  1-T_{n}\right)  /\sum_{n}T_{n}. \label{butt}
\end{equation}

However, we will not use the above expression (or its generalization to the non-linear case~\cite{buttiker}). 
Instead, we will derive an expression of shot noise in terms of single-particle 
wavefunctions.~\cite{chendiventra2003} This choice is a convenient one: we calculate the latter quantities 
directly from the integral form of the Schr\"odinger equation within the
density-functional theory of many-electron systems.~\cite{langt,diventralang021,diventralang022} We repeat here for the reader the derivation 
of such an expression~\cite{chendiventra2003} and we demonstrate that it reduces to Eq.~(\ref{butt}) when the transmission 
probabilities are extracted from the single-particle wavefunctions. 

We start by introducing the field operator of propagating electrons for a
sample connected to a left (L) and right (R) electrode in terms of
single-particle wavefunctions $\Psi_{E}^{L(R)}\left(  \mathbf{r,\mathbf{K}%
_{\Vert}}\right)  $ with energy $E$ (atomic units are used throughout this paper) and component of the momentum parallel to
the electrode surface $\mathbf{K}_{\parallel}$~\cite{langt,diventralang021,diventralang022} 
\begin{equation}
\hat{\Psi}=\hat{\Psi}^{L}+\hat{\Psi}^{R},
\end{equation}
where
\begin{equation}
\hat{\Psi}^{L(R)}=\sum_{E}a_{E}^{L(R)}\left(  t\right)  \Psi_{E}^{L(R)}\left(
\mathbf{r},\mathbf{K}_{\Vert}\right).~\label{field} 
\end{equation}
In this expression $a_{E}^{L(R)}\left(  t\right)  =\exp(-i\omega t)a_{E}^{L(R)}$, where  
$a_{E}^{L(R)}$ are the annihilation operators for electrons incident from the left (right)
electrode, satisfying the usual anticommutation relations: $%
\{a_{E}^{i\dag },a_{E^{\prime }}^{j}\}=\delta _{ij}\delta \left( E-E^{\prime
}\right) $. We assume that the electrons coming from the left (right) electrode 
thermalize completely far away from the sample, i.e., the statistics of electrons 
coming from left (right) is determined by the equilibrium 
Fermi-Dirac distribution function $f_{E}^{L\left(  R\right)  }$ deep into the left (right)
electrode, i.e.,
\begin{equation}
<a_{E}^{i\dag }a_{E^{\prime }}^{j}>=\delta _{ij}\delta \left( E-E^{\prime
}\right) f_{E}^{i},
\end{equation}%
with i, j=R, L.

In order to obtain the wave functions, $\Psi^{L(R)}_{E}(\mathbf{r},\mathbf{K_{\Vert}})$, we solve 
the Lippman-Schwinger equation self-consistently 

\begin{align}
& \Psi^{L(R)}_{E}(\mathbf{r},\mathbf{K_{\Vert}})    =\Psi^{L(R)}_{0,E,\mathbf{K_{\Vert}}}
(\mathbf{r}) \nonumber\\
& + \int d\mathbf{r}_{1}\int d\mathbf{r}_{2}G(\mathbf{r},\mathbf{r_{1}}) 
V(\mathbf{r_{1}},\mathbf{r_{2}})\Psi^{L(R)}_{E}(\mathbf{r_{2}},\mathbf{K_{\Vert}})
\text{.}
\end{align}

$\Psi^{L(R)}_{0,E,\mathbf{K_{\Vert}}}$ are the wave functions incident from the left (right) bare biased electrodes, and $G$ is the corresponding Green's function. $V$ is the difference between the potential of the 
total system and the one of the bare biased electrodes. It includes the sum of the nuclear, Hartree and exchange-correlation potentials.~\cite{langt,diventralang021,diventralang022}

>From the field operator Eq.~(\ref{field}) we can define the current operator
\begin{equation}
\hat{I}(z,t)=-i\int d\mathbf{R}\int d\mathbf{K}_{\Vert }\left( \hat{\Psi}%
^{\dag }\partial _{z}\hat{\Psi}-\partial _{z}\hat{\Psi}^{\dag }\hat{\Psi}
\right).
\end{equation}
The ensemble average current at zero temperature is thus 
\begin{equation}
<\hat{I}>=-i\int_{E_{FL}}^{E_{FR}}dE\int d\mathbf{R}\int d\mathbf{K}_{\Vert
}\left(  \tilde{I}_{E,E}^{R,R}\right) ,\label{average}%
\end{equation}
where
\begin{equation}
\tilde{I}_{E,E^{\prime}}^{ij}=\left(  \Psi_{E}^{i}\right)  ^{\ast}\nabla
\Psi_{E^{\prime}}^{j}-\nabla\left(  \Psi_{E}^{i}\right)  ^{\ast}%
\Psi_{E^{\prime}}^{j}\text{,} \label{cur}%
\end{equation}
with i, j=R, L. We have also assumed the right chemical potential to be higher than the left one.

The shot noise spectral density is defined as the Fourier transform of the current autocorrelation function
\begin{equation}
2\pi S\left(  \omega\right)  =\int dte^{i\omega t}\left\langle \Delta\hat
{I}\left(  t\right)  \Delta\hat{I}\left(  0\right)  \right\rangle,
\end{equation}
where $\Delta\hat{I}(t)$ is the excess current operator with respect to the average current defined above. 
Using the Bloch-De Dominicis theorem~\cite{bloch} we can express $S(\omega)$ as
\begin{align}
S\left(  \omega\right)   &  =\sum_{i,j=L,R}\int dEf_{E+\omega}^{i}\left(
1-f_{E}^{j}\right)  \cdot\nonumber\\
&  \int d\mathbf{R}_{1}\int d\mathbf{K}_{1}\tilde{I}_{E+\omega,E}^{ij}\int
d\mathbf{R}_{2}\int d\mathbf{K}_{2}\tilde{I}_{E,E+\omega}^{ji}\text{.} \label{eq7}
\end{align}
Taking the zero frequency and zero temperature limit, the shot noise, S, can finally be written as~\cite{chendiventra2003}
\begin{equation}
S=\int_{E_{FL}}^{E_{FR}}dE\left\vert \int d\mathbf{R}\int d\mathbf{K}\tilde
{I}_{E,E}^{LR}\right\vert ^{2}\text{,}
\label{noise}
\end{equation}
where $\tilde{I}_{E,E}^{LR}$ is defined in Eq.~(\ref{cur}). 

In order to demonstrate that Eq.~(\ref{noise}) reduces to Eq.~(\ref{butt}) we consider only a single 
one-dimensional scattering channel (generalization to many channels is straightforward). In this case the 
right- and left-moving waves satisfy the boundary conditions (with obvious notation)
\begin{equation}
\Psi^{R}_{E}=(2\pi)^{3/2}k_L^{-1/2} \times 
\left\{ \begin{array}{ll}
e^{ik_Lx}+R_{L}e^{-ik_Lx} & x<-\infty \\
T_{R}e^{ik_Rx} & x>\infty 
\end{array}\right.
\end{equation}
and
\begin{equation}
\Psi^{L}_{E}=(2\pi)^{3/2}k_R^{-1/2} \times 
\left\{ \begin{array}{ll}
T_{L}e^{-ik_Lx} & x<-\infty
\\e^{-ik_Rx}+R_{R}e^{ik_Rx} & x>\infty\text{,}
\end{array} \right.
\end{equation}
respectively. 

Introducing the above expressions into Eq.~(\ref{cur}) and re-writing Eq.~(\ref{noise}) as 
\begin{equation}
S(x_1,x_2)=\int_{E_{FL}}^{E_{FR}}dE \hat{I}^{RL}_{EE}(x_1)\hat{I}^{LR}_{EE}(x_2)\text{,} \label{1dshot}
\end{equation}
we can define $S^{ij}=S(x_1\to\pm\infty,x_2\to\pm\infty)$, 
with i, j=R, L. We thus obtain that (with $T_R=T_L=T$)
\begin{equation}
S^{LL}= \frac{V_0}{\pi}T(1-T),\label{sll}
\end{equation}
where we have implicitly assumed that $E_{FR}-E_{FL}=eV_0$, where $V_0$ is the external bias, and $T$ 
is independent of energy. For $T\ll 1$ we obtain the Poisson limit of shot noise 
$S^{P}= \frac{V_0}{\pi}T$. Taking the ratio of Eq.~(\ref{sll}) with $S^P$ leads to Eq.~(\ref{butt}).
Similarly we can prove that the following relations are valid~\cite{buttiker}
\begin{equation}
S^{LL}=S^{RR}=-S^{RL}=-S^{LR}\text{,}
\end{equation}
as a direct consequence of current conservation. 

As a test of Eq.~(\ref{noise}) we have calculated the 
noise for a single gold atom between two bulk electrodes modeled by jellium with $r_s= 3$. The gold 
atom is placed at 2.2 a.u. from each jellium edge. For a bias of 
$0.1V$ we obtain a conductance of $1.1 G_0$ ($G_0=2e^2/h$) and a Fano factor of 0.14. 
Experimental results on gold point contacts indicate that to such value of conductance corresponds a 
Fano factor of $0.16\pm0.03$,~\cite{van1,van2} indicating a good agreement between theory and experiment 
in the present case.

\section{Results and discussion}

\begin{figure}
\begin{center} 
\includegraphics[width=.48\textwidth]{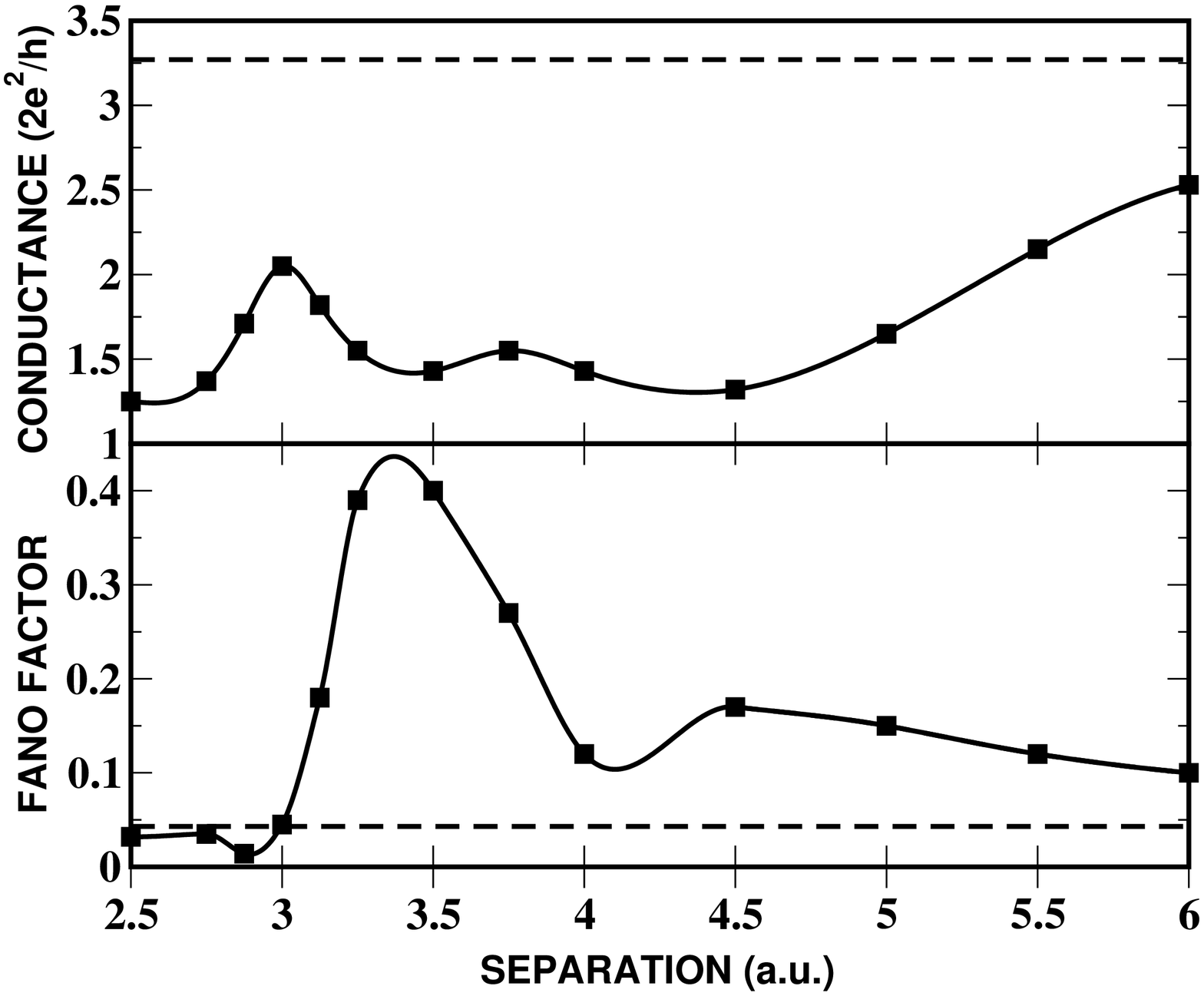}
\caption{Conductance (upper panel), in units of quantum of conductance $G_0$, and Fano factor (lower panel) 
as a function of wire separation for two parallel 4-atom wires. Filled squares 
correspond to calculated values. Horizontal broken lines correspond to twice 
the conductance of a single wire (upper panel) and the Fano factor of a single wire (lower panel).}
\label{fig2}
\end{center}
\end{figure}

\begin{figure}
\begin{center} 
\includegraphics[width=0.48\textwidth]{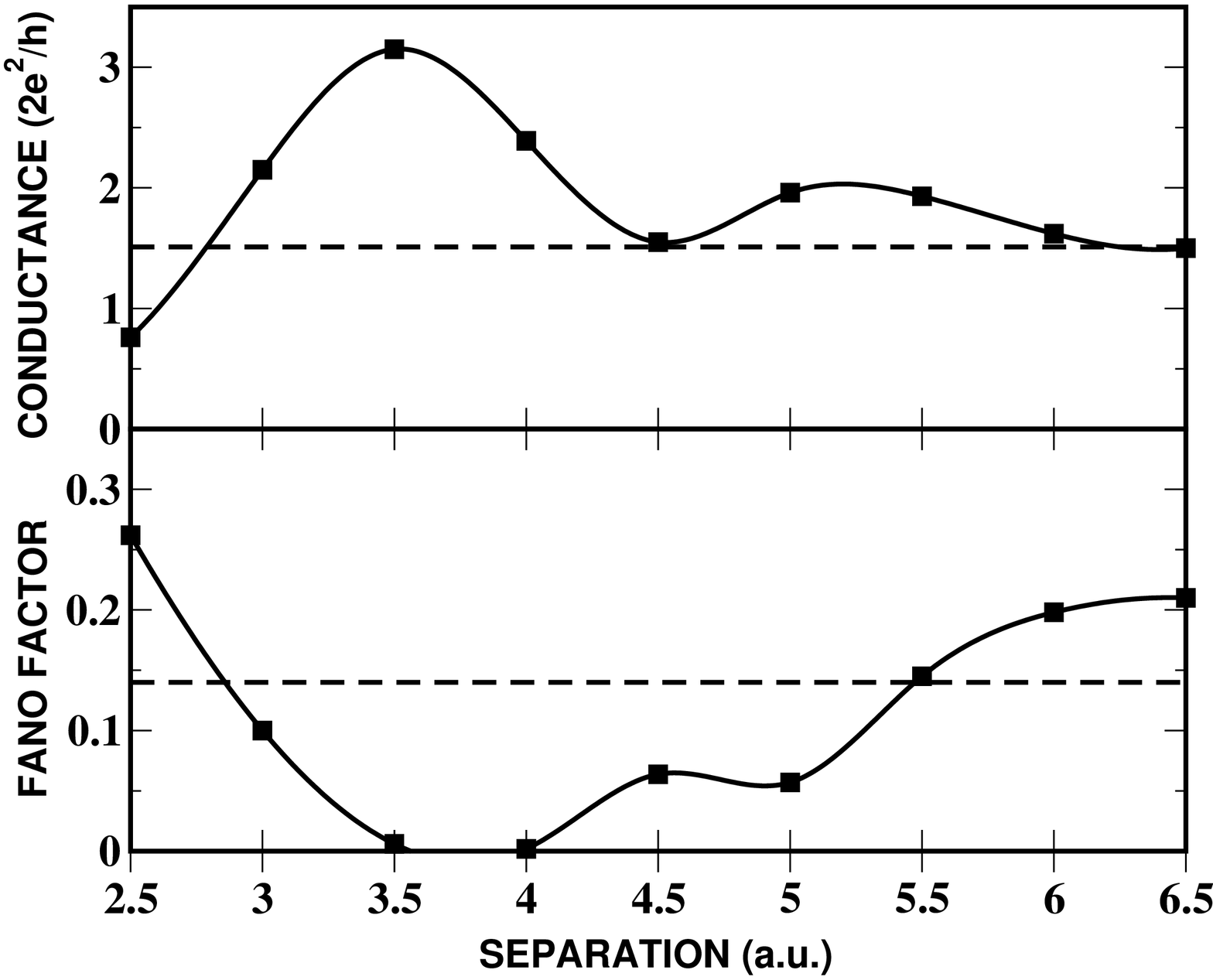}
\caption{Conductance (upper panel), in units of quantum of conductance $G_0$, and Fano factor (lower panel)
as a function of wire separation for two parallel 5-atom wires. Filled squares 
correspond to calculated values. Horizontal broken lines correspond to twice 
the conductance of a single wire (upper panel) and the Fano factor of a single wire (lower panel).}
\label{fig3}
\end{center}
\end{figure}

We are now ready to discuss the results for the parallel carbon wires. We will consider two different wire 
lengths:  4-atom chains and 5-atom chains. In each wire, the C-C distance has been fixed at 2.5 a.u. and, 
due to the construction of the jellium model,~\cite{langdiventracomment03} the end carbon atoms of the 
chains are 1.4 a.u. inside the positive background edge of jellium.~\cite{langavouris2000} 
The value of the external bias is 
$V=0.01V$. The conductance in units of $G_0$ (top) and Fano factor (bottom) for these two 
systems as a function of the interwire distance are plotted in Figs.~\ref{fig2} and~\ref{fig3}. In order to 
keep the same numerical error in the calculation of current and noise for 
all interwire distances, and due 
to computational limits, we could not evaluate these quantities for distances larger than about 
6.5 a.u. In these figures we report both the actual data points (filled squares) calculated from first principles and 
an interpolation in between the data points (solid line).  We also indicate 
in the figures the ``uncoupled-wire'' case for both the current (twice the current of each independent 
wire, i.e. without any wire-wire interaction) 
and Fano factor (equal to the Fano factor of each wire - the current doubles as well the noise, leaving the 
ratio unchanged). It is immediately evident from both figures that both the current and the Fano factor change 
in a non-linear way with interwire distance. However, the variation of noise is even more dramatic. We  
discuss the current first. 

It is clear from Figs.~\ref{fig2} and~\ref{fig3} that the conductance variation depends on the length of the 
carbon wires. For instance, the conductance reaches a value of about twice the conductance of each independent wire 
at about 6.5 a.u. for the 5-atom chain, while this is not the case for the 4-atom chain. In addition, the conductance of the system at infinite 
distance (i.e. in the absence of wire-wire interactions) is larger for the 4-atom wire than for the 5-atom 
wire. This difference is related to the fact that 
even-atom-number wires have fully occupied $\pi$ states, 
while the odd-atom-number wires have a half-filled $\pi$ state at the Fermi 
level.~\cite{langavouris2000,chendiventra2003} Note, however, that the band width 
of the metal electrodes (determined by the value of $r_s$) is much smaller in the present case than the 
one used in Ref.~\onlinecite{langavouris2000}. This implies that the extra $\sigma$ and $\pi$ states  
introduced by the coupling of the wires to the electrodes are truly bound states in our case, thus effectively reverting the trend of the change of conductance as a function of wire length: in the present case, 
the conductance (for large interwire separations) is larger for even number of carbon atoms than for 
odd number of carbon atoms.  Finally, the change in 
conductance is related to the detailed bonding properties of the parallel wires.~\cite{langavouris2000} 

\begin{figure}
\begin{center} 
\includegraphics[width=.48\textwidth]{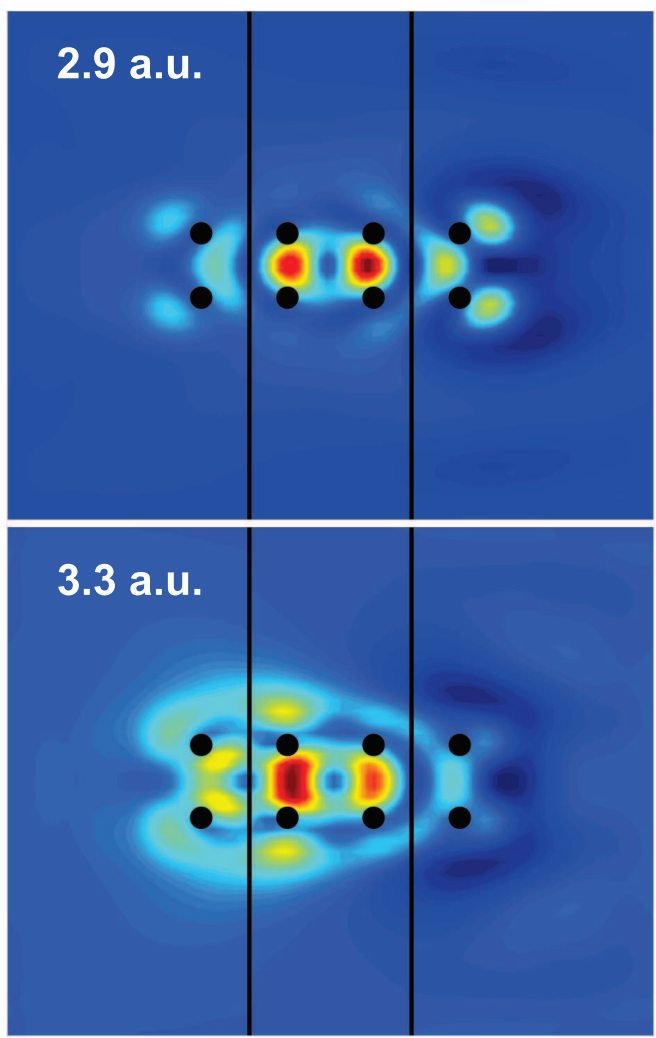}
\caption{Contour plot of the charge density for the 4-atom wire system at two different interwire distances. See text for the explicit definition of 
this density. Vertical black lines correspond to the edges of the jellium model and circles correspond to atomic 
positions.}
\label{fig4}
\end{center}
\end{figure}

It is evident in Fig.~\ref{fig2} and Fig.~\ref{fig3} that there is a correlation between the change in conductance as a 
function of interwire distance and the 
corresponding change in Fano factor. In general, whenever the conductance increases the Fano factor decreases, 
as can be understood intuitively from Eq.~(\ref{butt}). However, the Fano factor is more affected by the interwire interaction than the conductance.  We therefore discuss the bonding properties in connection 
to the Fano factor. In order to do so we show contour plots of the charge density. The charge density we consider is the one due to the populated global current-carrying 
states originating from the right electrode for the system with the
carbon atoms, minus the same quantity for the electrode-electrode system without the atoms. For the 4-carbon wire 
this quantity is plotted in Fig.~\ref{fig4} for a wire distance of about 2.9 a.u. (top) and 3.3 a.u. (bottom), i.e. 
for the case in which the noise is minimum (even lower than the value for the ``uncoupled-wire'' case) and maximum, respectively (see also Fig.~\ref{fig2}). It is evident from Fig.~\ref{fig4} (top) 
that at 2.9 a.u. the wires form two symmetric $\sigma$ bonds from each carbon $p$ orbital oriented parallel to the electrode surface and in the 
plane where the centers of all C atoms lie. These two $\sigma$ bonds link the four carbon atoms between the 
jellium edges as shown in Fig.~\ref{fig4} (top). In this configuration, electron waves travel from one electrode to the other as if the structure had two perfectly (degenerate) open channels 
(the value of the conductance in this case is also close to two, see Fig~\ref{fig2}). We stress however 
that this is only an analogy and does not necessarily imply that the number of open global channels 
in the structure is exactly two. On the 
other hand, at a distance of 3.3 a.u. (bottom of Fig.~\ref{fig4}), there is more redistribution of charge 
from the interchain $\sigma$ bonds to the $\pi$ bonds along each wire, thus increasing backscattering and 
the correlation between electron waves on each side of the structure. This feature is also present 
(even if at different distances) for the 5-atom chain. In this case, the formation of three quasi-degenerate 
$\sigma$ bonds extends from about 3.5 a.u. to about 3.7 a.u. giving rise to a conductance 
of about 3$G_0$ and a (almost) zero Fano factor in the same range (see Fig.~\ref{fig3}). 
The Fano factor is thus very sensitive to 
wire-wire interactions, and in particular to the charge redistribution between bonds. Since even small charge redistributions can increase or reduce backscattering, the Fano factor can change by simply changing the 
interwire distance by less than 0.5\AA. 

\begin{figure}
\begin{center} 
\includegraphics[width=.48\textwidth]{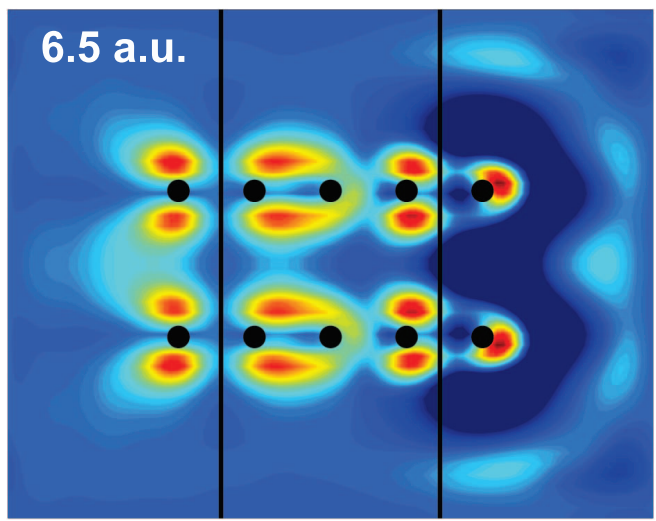}
\caption{Contour plot of the charge density for the 5-atom wire system at interwire distance of 6.5 a.u. See text for the explicit definition of this 
density. Vertical black lines correspond to the edges of the jellium model and circles correspond to atomic 
positions.}
\label{fig5}
\end{center}
\end{figure}

When the two wires are separated by a relatively large distance, the wire-wire interaction is negligible and the 
current reaches a value close to twice the value of the current of each wire. However, correlations between 
waves travelling in the two different wires do not necessarily go to zero, and this would affect the noise. 
This can be understood by 
looking at the contour plot of the charge density (as defined above) for a 5-C wire at 
interwire distance of 6.5 a.u. (Fig.~\ref{fig5}). At this distance the conductance is almost exactly 
twice the value of the conductance of each wire (see Fig.~\ref{fig2}). However, the Fano factor is still 
{\it larger} than the Fano factor of each wire. This is because, even though there is essentially 
no bonding between the carbon atoms of the two wires (see Fig.~\ref{fig5}), 
there is still some probability for a wave that gets out of a wire on one side of the junction to travel 
{\it along} the surface of the electrode and get reflected back into the other wire from the same side of the junction. In other words, 
the global current-carrying wavefunctions have finite extension in the surface region between the wires. 
Indeed, for large wire separations, the lateral component of the scattering 
wavefunctions resembles closely a Bloch wave within few atomic units from the surface region 
in between the wires. This wavefunction extension thus generates extra noise compared to the 
``uncoupled-wire'' case. An alternative and intuitive way of thinking about this point is the following: 
at large wire separations, the different paths along which an electron is carried from one lead to the other across either 
one of the two single wires, have energies that are quasi-degenerate. The system thus fluctuates fast between 
these quasi-degenerate states and additional noise is generated (in a manner similar to 
the generation of telegraph noise). 
Clearly, such correlations would be destroyed if 
inelastic effects were into play. However, as explained above, for the distances considered here 
these effects are less likely to contribute.  

We conclude by stressing that, as the present work also shows, 
shot noise is a very sensitive tool to probe electron transport 
properties. Therefore, in order to increase our understanding of transport in atomic-scale structures, 
it would be very desirable to have experimental results of noise properties of molecular wires. We hope 
our work will inspire future studies in this direction. However, we note 
that the extreme sensitivity of noise to bonding properties between parallel wires can actually 
constitute a limitation in electronic applications of nanoscale structures.

{\bf Acknowledgments} We thank M. B\"{u}ttiker for useful discussions. 
We acknowledge support from the NSF Grant Nos. DMR-01-02277 and
DMR-01-33075, and Carilion Biomedical Institute. Acknowledgement is also
made to the Donors of The Petroleum Research Fund, administered by the
American Chemical Society, for partial support of this research.

\newpage

\end{document}